\documentclass[conference,letterpaper]{IEEEtran}

\usepackage[utf8]{inputenc} 
\usepackage[T1]{fontenc}
\usepackage{url}
\usepackage{ifthen}
\usepackage{cite}
\usepackage{amssymb}
\usepackage{amsthm}
\usepackage{amsfonts}
\usepackage{graphicx}
\usepackage{algorithmic}
\usepackage[cmex10]{amsmath} 
\theoremstyle{definition}
\newtheorem{remk}{\textbf{Remark}}

\begin{document}
\title{Private Secure Coded Computation} 


\author{%
  \IEEEauthorblockN{Minchul Kim and Jungwoo Lee}
  \IEEEauthorblockA{Seoul National University\\
                    Department of Electrical and Computer Engineering\\
                    08826 Seoul, Korea\\
                    Email: kmc1222@cml.snu.ac.kr, junglee@snu.ac.kr}
}


\maketitle

\begin{abstract}

We introduce a variation of coded computation that ensures data security and master's privacy against workers, which is referred to as \textit{private secure coded computation}.
In private secure coded computation, the master needs to compute a function of its own dataset and one of the datasets in a library exclusively shared by external workers.
After recovering the desired function, the workers must not be able to know which dataset in the library was desired by the master or obtain any information about the master's own data. 
We propose a private secure coded computation scheme for matrix multiplication, namely  \textit{private secure polynomial codes}, based on \textit{private polynomial codes} for \textit{private coded computation}.
In simulations, we show that the private secure polynomial codes achieves better computation time than private polynomial codes modified for private secure coded computation.
\end{abstract}


\section{Introduction}
\label{intro}

In a distributed computing system where a master partitions a massive computation into smaller sub-computations and distributes these sub-computations to several workers in order to reduce the runtime to complete the whole computation, some slow workers can be bottleneck of the process. 
These slow workers are called \textit{stragglers}, and mitigating the effect of these stragglers is one of the major issues in distributed computing. 
Recently, a coding technique was introduced for straggler mitigation \cite{KLee}. 
In \cite{KLee}, for a matrix-vector multiplication, the matrix is $(n,k)$-MDS coded and distributed to $n$ workers so that each encoded matrix is assigned to one worker. 
Each worker multiplies the coded submatrix by a vector and returns the multiplication to the master. 
After $k$ out of $n$ workers return their multiplications, the master can recover the whole computation. 
Since the computation of the slowest $n-k$ workers is ignored, at most $n-k$ stragglers can be mitigated. 
This kind of approach to distributed computing is referred to as \textit{coded computation}. 
Several follow-up studies of coded computation were proposed \cite{followup1} -\cite{followup5}. 

In this paper, we introduce a variation of coded computation that considers both master's privacy and data security against the workers, which is referred to as \textit{private secure coded computation}.
In the private secure coded computation, the master requires distributed computing on a function $f$ of its own data $\mathbf{A}$ and specific data $\mathbf{B}_D$ included in a library $\mathbf{B}$, which is exclusively shared by external workers.
For each worker, the master encodes $\mathbf{A}$ with an encoding function $g^D_{\mathbf{A}}$, sends encoded data to the worker, and requests the worker to encode $\mathbf{B}$ with an encoding function $g^D_{\mathbf{B}}$ and compute a function $f_{\text{W}}(g^D_{\mathbf{A}}(\mathbf{A}),g^D_{\mathbf{B}}(\mathbf{B}))$. 
After the master recovers the result of desired function $f(\mathbf{A},\mathbf{B}_D)$ from the computation results of $f_{\text{W}}$ returned by the workers, the workers should not be able to identify that $\mathbf{B}_D$ is desired by the master, which would imply that the master's privacy is protected.
The workers also should not obtain any information about the master's data $\mathbf{A}$, which would imply that the data security is guaranteed.
Private secure coded computation will be explained in further detail in Section \ref{system}.

As a motivating example of the private secure coded computation, we may consider a user who employs an artificial intelligence (AI) assistant, e.g. Google Assistant or Siri, with its mobile.
We assume that the user can request a recommendation from an AI assistant of an item which is included in one of $M$ categories that the AI assistant can recommend, e.g. movies, games, restaurants, and so on.
We refer to the $M$ categories as a library $\mathbf{B}$ and denote them by $\{\mathbf{B}_k\}_{k=1}^M$ such that $\mathbf{B}=\{\mathbf{B}_k\}_{k=1}^M$.  
We also assume that the user stores its preference parameter $\mathbf{A}$.
When the user requests a recommendation from the AI assistant of an item in a category $\mathbf{B}_D$, the assistant encodes $\mathbf{A}$ and sends encoded data to several distributed workers, e.g. data centers, for recovering $f(\mathbf{A},\mathbf{B}_D)$ in a distributed way.
After recovery, the AI assistant can decide the recommended item based on $f(\mathbf{A},\mathbf{B}_D)$.
We assume that the AI assistant does not share the user preference parameter $\mathbf{A}$ with workers and that the user can delete the recommendation service usage record right after the item is recommended so that the AI assistant does not identify the user's recommendation service usage pattern.

In this example, the data security of $\mathbf{A}$ against the workers is ensured by encrypting $\mathbf{A}$ while encoding it. 
However, encrypting $\mathbf{A}$ cannot protect the user's privacy. 
Generally, the user uses this recommendation service according to its life cycle.
That is, if the workers track the recommendation service usage records, the user's life cycle is revealed to them, which implies that the user's privacy has been compromised.
We remark that this privacy invasion on the user's life cycle is related to $\mathbf{B}$, not $\mathbf{A}$.
Therefore, encrypting $\mathbf{A}$ cannot protect the user's privacy on the life cycle.
In order to protect the user's privacy, the workers should not know that a particular $\mathbf{B}_D$ is desired by a user, which motivates the private secure coded computation.

Data security in coded computation was studied in previous works \cite{Secure} -\cite{Secure3}.
In these works, the master has both of two matrices $\mathbf{A}$ and $\mathbf{B}_D$ whereas the workers do not have any library.
The master wants to compute a matrix multiplication $\mathbf{AB}_D$ using the workers.
Since the workers do not have their own library, the master's privacy against the workers is not considered in this system model.
The master's privacy in coded computation was considered in \cite{Arxiv} first, and the coded computation model that considers the master's privacy was referred to as \textit{private coded computation}.
In \cite{Arxiv}, an achievable scheme for private coded computation based on \textit{polynomial codes}\cite{PC} was proposed, which was referred to as \textit{private polynomial codes}.

In this paper, we propose a private secure coded computation scheme for matrix multiplication, based on polynomial codes.
We refer to this scheme as \textit{private secure polynomial codes}.
For the data security, the master jointly encode $\mathbf{A}$ and a random matrix $\mathbf{R}$ into polynomial codes where the random matrix $\mathbf{R}$ is exclusively owned by the master and concealed to the workers, which was previously proposed in \cite{Secure}.
The idea for protecting the master's privacy is based on private polynomial codes in \cite{Arxiv}.
In simulation results, we show that the private secure polynomial codes achieve faster computation time than private polynomial codes modified for private secure coded computation.

\textit{Notation }: We use $[N]$ to denote a set comprised of $N$ elements, 1 to $N$. A set comprised of $M$ elements, $N+1$ to $N+M$ is denoted by $[N+1:N+M]$.




\section{System model}
\label{system}

In this section, we describe a system model of private secure coded computation.
There is a master who has its own dataset $\mathbf{A}$, where $\mathbf{A}$ is an element (matrix) in a vector space $\mathbb{V}_1$ over a field $\mathbb{F}$.
There are also $N$ external workers $\{\text{W}_i\}_{i=1}^N$, and these workers share a library $\mathbf{B}$ which  consists of $M$ different datasets $\{\mathbf{B}_k\}_{k=1}^M$.
Each dataset $\mathbf{B}_k$ is an element (matrix) in a vector space $\mathbb{V}_2$ over the same field $\mathbb{F}$.
The master needs distributed computing on a function $f$ of $\mathbf{A}$ and one of $M$ datasets $\{\mathbf{B}_k\}_{k=1}^M$ in library $\mathbf{B}$, where $f:(\mathbb{V}_1,\mathbb{V}_2) \rightarrow \mathbb{V}_3$ for a vector space $\mathbb{V}_3$ over the same field $\mathbb{F}$.
We denote the desired dataset by $\mathbf{B}_D$.
Therefore, the whole computation desired by the master is denoted by $f(\mathbf{A},\mathbf{B}_D)$. 
Since we consider private coded computation for matrix multiplication, $f(\mathbf{A},\mathbf{B}_D)=\mathbf{AB}_D$ in this paper.

The whole computation is converted into several sub-computations and assigned to the workers.
Each worker returns its sub-computation result to the master.
When sufficient number of sub-computation results are returned to the master, the master can recover the whole computation $f(\mathbf{A},\mathbf{B}_D)$ based on the received sub-computation results. 
We denote the minimum number of sub-computation results to recover $f(\mathbf{A},\mathbf{B}_D)$ by $K$ which was referred to as \textit{recovery threshold} in \cite{PC}.
The slowest $N-K$ workers become stragglers, since they do not return their sub-computations results.
After the master recovers the whole computation, each worker should not be able to obtain any information about $\mathbf{A}$ or identify that $\mathbf{B}_D$ is desired by the master, thus ensuring data security and master's privacy.
In this paper, we assume that the workers do not collude with each other so that each worker does not know which sub-computations are assigned to, computed by, and returned by the other workers.

The master's own dataset $\mathbf{A}$ and the library $\mathbf{B}$ are encoded for the private secure coded computation.
Note that the master's own dataset $\mathbf{A}$ is encoded by the master whereas the library $\mathbf{B}$ is encoded by each worker.
The master encodes $\mathbf{A}$ for each worker $\text{W}_i$.
We denote the encoding function of $\mathbf{A}$ for the worker $\text{W}_i$ and desired matrix $\mathbf{B}_D$ by $g^D_{\mathbf{A},\text{W}_i}$, where $g^D_{\mathbf{A},\text{W}_i}:\mathbb{V}_1 \rightarrow \mathbb{U}_1$ for a vector space $\mathbb{U}_1$ over the same field $\mathbb{F}$.
The master sends the encoded data $g^D_{\mathbf{A},\text{W}_i}(\mathbf{A})$ to the worker $\text{W}_i$ and also sends the queries for requesting $\text{W}_i$ to encode the library $\mathbf{B}$.
We denote the encoding function of the worker $\text{W}_i$ for the library $\mathbf{B}=\{\mathbf{B}_k\}_{k=1}^M$ and the desired dataset $\mathbf{B}_D$ by $g^D_{\mathbf{B},\text{W}_i}$, where $g^D_{\mathbf{B},\text{W}_i}:\mathbb{V}_2^M \rightarrow \mathbb{U}_2$ for a vector space $\mathbb{U}_2$ over the same field $\mathbb{F}$.
The master also sends the queries to the worker $\text{W}_i$ to compute a function of $g^D_{\mathbf{A},\text{W}_i}(\mathbf{A})$ and $g^D_{\mathbf{B},\text{W}_i}(\mathbf{B})$ and return the computation result of the function to the master.
That is, the worker $\text{W}_i$ computes $f^D_{\text{W}_i}(g^D_{\mathbf{A},\text{W}_i}(\mathbf{A}), g^D_{\mathbf{B},\text{W}_i}(\mathbf{B}))$.
We denote the function of $\text{W}_i$ by $f^D_{\text{W}_i}:(\mathbb{U}_1,\mathbb{U}_2)\rightarrow \mathbb{U}_3$ for a vector space $\mathbb{U}_3$ over the same field $\mathbb{F}$.
Without considering where the sub-computation result comes from, we denote the $i$th sub-computation result returned to the master by $S_i$, where $S_i$ is an element in the vector space $\mathbb{U}_3$. 
After $K$ sub-computation results $\{S_i\}_{i=1}^K$ across the $N$ workers are returned to the master, the master can recover the whole computation $f(\mathbf{A},\mathbf{B}_D)$ by decoding $\{S_i\}_{i=1}^K$.
If we denote the decoding function at the master by $d_D:\mathbb{U}_3^K\rightarrow \mathbb{V}_3$, the decoding function $d_D$ should satisfy the constraint given by $d_D(S_1, S_2, \cdots , S_K)=f(\mathbf{A},\mathbf{B}_D)$.

The master's privacy is protected when none of the workers can identify index $D$ of the desired dataset $\mathbf{B}_D$ after the master recovers the whole computation. Since the privacy we consider is information-theoretic privacy, the privacy constraint for each worker $\text{W}_i$ can be expressed as 
\begin{align}
 I(D;Q_i^D,g^D_{\mathbf{A},\text{W}_i}(\mathbf{A}),f^D_{\text{W}_i}(g^D_{\mathbf{A},\text{W}_i}(\mathbf{A}), g^D_{\mathbf{B},\text{W}_i}(\mathbf{B})),\mathbf{B})=0, \nonumber
\end{align}
where $Q_i^D$ denotes the queries that the master sends to the worker $\text{W}_i$ for encoding $g^D_{\mathbf{B},\text{W}_i}(\mathbf{B})$ and computing $f^D_{\text{W}_i}(g^D_{\mathbf{A},\text{W}_i}(\mathbf{A}), g^D_{\mathbf{B},\text{W}_i}(\mathbf{B}))$.

For a simpler expression, we denote $g^D_{\mathbf{A},\text{W}_i}(\mathbf{A})$ and $f^D_{\text{W}_i}(g^D_{\mathbf{A},\text{W}_i}(\mathbf{A}), g^D_{\mathbf{B},\text{W}_i}(\mathbf{B}))$ by $C_i^D$ and $R_i^D$, respectively, so that the privacy constraint becomes
\begin{align}
 I(D;Q_i^D,C_i^D,R_i^D,\mathbf{B})=0. 
\label{privacyconstraint}
\end{align}

Similarly, the data security constraint for each worker $\text{W}_i$ can be expressed as 
\begin{align}
 I(\mathbf{A};Q_i^D,C_i^D,R_i^D,\mathbf{B})=0.. 
\label{securityconstraint}
\end{align}

\begin{figure}
    \centering
    \includegraphics[width=8.5cm]{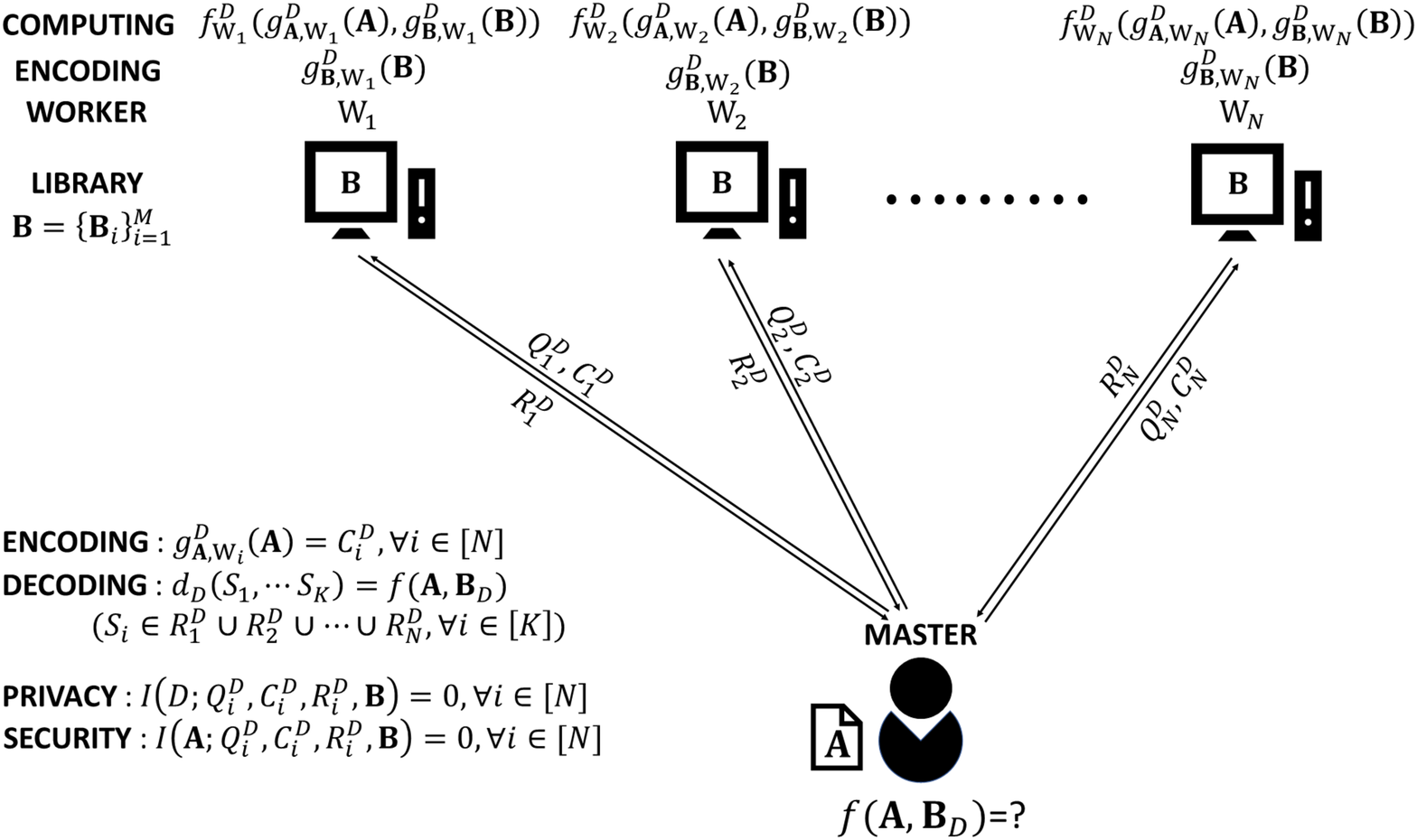}
    \caption{
    The overall process of private secure coded computation. 
        }
    \label{overall}
\end{figure}

The overall process of the private secure coded computation is depicted in Fig. \ref{overall}.

\section{Private Secure Polynomial Codes}
\label{pspc}

In this section, we propose private secure polynomial codes for matrix multiplication.
We describe the scheme with an illustrative example and generally describe the private secure polynomial codes. 
We also prove that the master's privacy and data security are protected.

\subsection{Illustrative Example}
\label{example}

We assume that the master has a matrix 
$\mathbf{A}\in \mathbb{F}_q^{r\times s}$ 
for sufficiently large finite field 
$\mathbb{F}_q$ 
and that there are 12 non-colluding workers 
$\{\text{W}_{n}\}_{n=1}^{12}$ 
where each worker has a library of two matrices 
$\mathbf{B}_{1},\mathbf{B}_{2}\in \mathbb{F}_q^{s\times t}$.
As in Section \ref{system}, we denote the library by $\mathbf{B}$.
Let us assume that the master wants to compute 
$\mathbf{AB}_1$ 
using 
$\{\text{W}_{n}\}_{n=1}^{12}$
while hiding that the master desires 
$\mathbf{B}_{1}$ 
from the workers. 
The matrix $\mathbf{A}$ can be partitioned into two submatrices 
$\mathbf{A}_{0}, \mathbf{A}_{1}\in \mathbb{F}_q^{r/2\times s}$ 
so that 
$\mathbf{A}$=$\begin{bmatrix}
{\mathbf{A}_0} \\ {\mathbf{A}_1}
\end{bmatrix}$ and each of $\mathbf{B}_1,\mathbf{B}_2$ 
are partitioned into two submatrices 
$\mathbf{B}_{k,1}, \mathbf{B}_{k,2}\in \mathbb{F}_q^{s\times t/2}, k\in[2]$
, so that 
$\mathbf{B}_k$=$\begin{bmatrix}
{\mathbf{B}_{k,1}} & {\mathbf{B}_{k,2}}
\end{bmatrix}$. 
Therefore, 
$\mathbf{AB}_1=\begin{bmatrix}
\mathbf{A}_0\mathbf{B}_{1,1} & \mathbf{A}_0\mathbf{B}_{1,2} \\ \mathbf{A}_1\mathbf{B}_{1,1} & \mathbf{A}_1\mathbf{B}_{1,2}
\end{bmatrix}$. 
The private secure polynomial codes for $\mathbf{A}$,  
$\mathbf{B}_{1}$ and $\mathbf{B}_{2}$
are as follows.
\begin{gather}
\mathbf{\tilde A}(x)=\mathbf{A}_0+\mathbf{A}_1 x+\mathbf{R}x^2, \:\:\:\: \mathbf{\tilde B}_k(x)=\mathbf{B}_{k,1} x^3+\mathbf{B}_{k,2} x^{6}, \nonumber
\end{gather}
where $k\in[2]$, $\mathbf{R}\in \mathbb{F}_q^{r/2 \times s}$ denotes a random matrix, and $x \in \mathbb{F}_q$ denotes the variable of polynomials $\mathbf{\tilde A}$ and $\mathbf{\tilde B}_k$. 

We denote the evaluations of 
$\mathbf{\tilde A}$ and $\mathbf{\tilde B}_k$ at $x=x_i$ by 
$\mathbf{\tilde A}(x_i)$ and $\mathbf{\tilde B}_k(x_i)$
, respectively. 
For the desired matrix $\mathbf{B}_1$ and each worker $\text{W}_i$, the master evaluates $\mathbf{\tilde A}$ at a randomly chosen point $x_{i}$ and sends the evaluation $\mathbf{\tilde A}(x_{i})$ to the worker $\text{W}_i$.
That is, $g^1_{\mathbf{A},\text{W}_i}(\mathbf{A})=\mathbf{\tilde A}(x_{i})$. 
We assume that the points $\{x_{i}\}_{i=1}^{12}$ are distinct from each other.
The master also sends the queries $Q_i^1$ that request $\text{W}_i$ to encode the library $\mathbf{B}$ with an encoding function $g^1_{\mathbf{B},\text{W}_i}$ and compute a function $f_{\text{W}_i}^1(g^1_{\mathbf{A},\text{W}_i}(\mathbf{A}),g^1_{\mathbf{B},\text{W}_i}(\mathbf{B}))$.

The library $\mathbf{B}$ is encoded as follows.
Firstly, for each worker $\text{W}_i$, $\mathbf{\tilde B}_1$ is evaluated at $x_i$. 
Secondly, for all of the workers, the undesired matrix $\mathbf{\tilde B}_2$ is evaluated at a randomly chosen point $x_{13}$ which is distinct from the points $\{x_i\}_{i=1}^{12}$.
Since the workers do not collude with each other, they cannot notice that  $\mathbf{\tilde B}_2$ is evaluated at an identical point $x_{13}$ across workers.
Finally, for each worker $\text{W}_i$, the encoded library is given by $g^1_{\mathbf{B},\text{W}_i}(\mathbf{B})=\mathbf{\tilde B}_1(x_{i})+\mathbf{\tilde B}_2(x_{13})$.

After encoding the library, each worker $\text{W}_i$ computes a function $f_{\text{W}_i}^1(g^1_{\mathbf{A},\text{W}_i}(\mathbf{A}),g^1_{\mathbf{B},\text{W}_i}(\mathbf{B}))=\mathbf{\tilde A}(x_{i})(\mathbf{\tilde B}_1(x_i)+\mathbf{\tilde B}_2(x_{13}))$ which is given by
\begin{flalign}
&\mathbf{\tilde A}(x_{i})(\mathbf{\tilde B}_1(x_i)+\mathbf{\tilde B}_2(x_{13}))\nonumber&& \\ &=(\mathbf{A}_0+\mathbf{A}_1x_{i}+\mathbf{R}x_i^2) \times \nonumber && \\ 
&\:\:\:\:\:\:(\mathbf{B}_{1,1} x_i^3+\mathbf{B}_{1,2} x_i^{6}+\mathbf{B}_{2,1} x_{13}^3+\mathbf{B}_{2,2} x_{13}^{6})&& \nonumber \\ \nonumber &=\sum_{l=0}^{8}{\mathbf{Z}_{l} x_{i}^l},&&  \nonumber 
\end{flalign}
where $\{\mathbf{Z}_{l}\}_{l=0}^{8}$ are given by
\begin{gather}
\mathbf{Z}_{l}=\mathbf{A}_{l}(\mathbf{B}_{2,1} x_{13}^3+\mathbf{B}_{2,2} x_{13}^{6}) \:\:\: \forall l\in[0:1], \nonumber \\ 
\mathbf{Z}_{l}=\mathbf{A}_{l-3}\mathbf{B}_{1,1} \:\:\: \forall l\in[3:4], \nonumber \\ 
\mathbf{Z}_{l}=\mathbf{A}_{l-6}\mathbf{B}_{1,2} \:\:\: \forall l\in[6:7],  \nonumber \\ 
\mathbf{Z}_{2}=\mathbf{R}(\mathbf{B}_{2,1} x_{13}^3+\mathbf{B}_{2,2} x_{13}^{6}), \mathbf{Z}_{5}=\mathbf{R}\mathbf{B}_{1,1}, \mathbf{Z}_{8}=\mathbf{R}\mathbf{B}_{1,2}. \nonumber 
\end{gather}

Since the degree of polynomial $\mathbf{\tilde A}(x)(\mathbf{\tilde B}_1(x)+\mathbf{\tilde B}_2(x_{13}))$ is 8 and the evaluating points $\{x_{i}\}_{i=1}^{12}$ are distinct from each other, the master can decode the polynomial from the sub-computation results returned by the 9 fastest workers, by polynomial interpolation.
We denote the decoding function by $d_D$ and sub-computation result returned from the $i$th fastest worker by $S_{i}$.
The master can decode the polynomial $\mathbf{\tilde A}(x)(\mathbf{\tilde B}_1(x)+\mathbf{\tilde B}_2(x_{13}))$ from $\{S_{i}\}_{i=1}^9$, so that the coefficients $\{\mathbf{Z}_{l}\}_{l=0}^{8}$ are obtained.
Note that the term $\mathbf{Z}_0=\mathbf{A}_0\mathbf{B}_{2,1} x_{13}^3+\mathbf{A}_0\mathbf{B}_{2,2} x_{13}^6$ and $\mathbf{Z}_1=\mathbf{A}_1\mathbf{B}_{2,1} x_{13}^3+\mathbf{A}_1\mathbf{B}_{2,2} x_{13}^6$ are constant terms in each polynomial, respectively.
The whole computation $\mathbf{AB}_1=\begin{bmatrix}
\mathbf{A}_0\mathbf{B}_{1,1} & \mathbf{A}_0\mathbf{B}_{1,2} \\ \mathbf{A}_1\mathbf{B}_{1,1} & \mathbf{A}_1\mathbf{B}_{1,2}
\end{bmatrix}$ can be recovered from the coefficients of $x^3$,$x^4$,$x^6$,$x^7$.
Therefore, the recovery threshold $K$ equals 9 and  $d_D(S_{1}, S_{2}, \cdots, S_{9})=\mathbf{AB}_1$.

\subsection{General Description}
\label{general}
In this section, we generally describe the private secure polynomial codes for matrix multiplication.
There are $N$ non-colluding workers $\{\text{W}_{n}\}_{n=1}^{N}$ and each worker has a library $\mathbf{B}$ of $M$ matrices $\{\mathbf{B}_{k}\}_{k=1}^{M}$ where each $\mathbf{B}_k\in  \mathbb{F}_q^{s\times t}$ for sufficiently large finite field $\mathbb{F}_q$. 
The master has a matrix $\mathbf{A}\in \mathbb{F}_q^{r\times s}$ and desires to multiply $\mathbf{A}$ by one of $\{\mathbf{B}_{k}\}_{k=1}^{M}$ in the library $\mathbf{B}$ while keeping the index of desired matrix $\mathbf{B}_{D}$ and content of $\mathbf{A}$ from all of the workers. 
Matrix $\mathbf{A}$ can be partitioned into $m$ submatrices $\{\mathbf{A}_{k}\}_{k=0}^{m-1}\in \mathbb{F}_q^{r/m\times s}$ and each $\mathbf{B}_k$ can be partitioned into $n-1$ submatrices $\{\mathbf{B}_{k,l}\}_{l=1}^{n-1}\in \mathbb{F}_q^{s\times t/(n-1)}$, where $m,n\in \mathbb{N}^+$. 
The whole computation $\mathbf{AB}_D$ that the master wants to recover is given by
\begin{align}
\mathbf{AB}_D=\begin{bmatrix}
\mathbf{A}_0\mathbf{B}_{D,1} & \mathbf{A}_0\mathbf{B}_{D,2} & \cdots & \mathbf{A}_0\mathbf{B}_{D,n-1}  \\ \mathbf{A}_1\mathbf{B}_{D,1} & \mathbf{A}_1\mathbf{B}_{D,2} & \cdots & \mathbf{A}_1\mathbf{B}_{D,n-1} \\
\vdots & \vdots & \cdots & \vdots \\
\mathbf{A}_{m-1}\mathbf{B}_{D,1} & \mathbf{A}_{m-1}\mathbf{B}_{D,2} & \cdots & \mathbf{A}_{m-1}\mathbf{B}_{D,n-1} 
\end{bmatrix}.
\nonumber
\end{align}

The polynomial codes for $\mathbf{A}$ and $\{\mathbf{B}_{k}\}_{k=1}^M$ are given as follows.
\begin{align}
\mathbf{\tilde A}(x)=\sum_{l=0}^{m-1}\mathbf{A}_l x^{l}+\mathbf{R}x^m, \:\:\:\: \mathbf{\tilde B}_k(x)=\sum_{l=1}^{n-1}\mathbf{B}_{k,l} x^{l(m+1)}, \nonumber
\label{encoding}
\end{align}
where $k\in[M]$, $\mathbf{R}\in \mathbb{F}_q^{r/m \times s}$ denotes a random matrix, and $x \in \mathbb{F}_q$ denotes the variable of polynomials $\mathbf{\tilde A}$ and $\mathbf{\tilde B}_k$.

We denote the evaluations of 
$\mathbf{\tilde A}$ and $\mathbf{\tilde B}_k$ at $x=x_i$ by 
$\mathbf{\tilde A}(x_i)$ and $\mathbf{\tilde B}_k(x_i)$, respectively. 
For the desired matrix $\mathbf{B}_D$ and each worker $\text{W}_i$, the master evaluates $\mathbf{\tilde A}$ at a randomly chosen point $x_{i}$ and sends the evaluation $\mathbf{\tilde A}(x_{i})$ to the worker $\text{W}_i$.
That is, $g^D_{\mathbf{A},\text{W}_i}(\mathbf{A})=\mathbf{\tilde A}(x_{i})$. 
We assume that the points $\{x_{i}\}_{i=1}^{N}$ are distinct from each other.
The master also sends the queries $Q_i^D$ that request $\text{W}_i$ to encode the library $\mathbf{B}$ with an encoding function $g^D_{\mathbf{B},\text{W}_i}$ and compute a function $f_{\text{W}_i}^D(g^D_{\mathbf{A},\text{W}_i}(\mathbf{A}),g^D_{\mathbf{B},\text{W}_i}(\mathbf{B}))$.

The library $\mathbf{B}$ is encoded as follows.
Firstly, for each worker $\text{W}_i$, $\mathbf{\tilde B}_D$ is evaluated at $x_i$. 
Secondly, for all of the workers and the undesired matrices $\{\mathbf{\tilde B}_k|k\in[M]\setminus D\}$, each undesired matrix $\mathbf{\tilde B}_k$ is evaluated at a randomly chosen point $x_{j_k}$ which is distinct from the points $\{x_i\}_{i=1}^{N}$.
Since the workers do not collude with each other, they cannot notice that each $\mathbf{\tilde B}_k$ is evaluated at an identical point $x_{j_k}$ across workers.
Finally, for each worker $\text{W}_i$, the encoded library is given by $g^D_{\mathbf{B},\text{W}_i}(\mathbf{B})=\mathbf{\tilde B}_D(x_{i})+\sum_{k\in [M]\setminus D}{\mathbf{\tilde B}_k(x_{j_k})}$.

After encoding the library, each worker $\text{W}_i$ computes a function $f_{\text{W}_i}^D(g^D_{\mathbf{A},\text{W}_i}(\mathbf{A}),g^D_{\mathbf{B},\text{W}_i}(\mathbf{B}))=\mathbf{\tilde A}(x_{i})(\mathbf{\tilde B}_D(x_i)+\sum_{k\in [M]\setminus D}{\mathbf{\tilde B}_k(x_{j_k})})$ which is given by
\begin{flalign}
&\mathbf{\tilde A}(x_{i})(\mathbf{\tilde B}_D(x_i)+\sum_{k\in [M]\setminus D}{\mathbf{\tilde B}_k(x_{j_k})})\nonumber&& \\ &=\mathbf{\tilde A}(x_i)\mathbf{\tilde B}_D(x_i)+\mathbf{\tilde A}(x_i)\sum_{k\in[M]\setminus D}{\mathbf{\tilde B}_k(x_{j_k})}&& \nonumber \\
&=\sum_{l=0}^{m-1}\sum_{p=1}^{n-1}\mathbf{A}_{l}\mathbf{B}_{D,p}x^{l+p(m+1)}+\sum_{p=1}^{n-1}\mathbf{R}\mathbf{B}_{D,p}x^{pm+m+p}&& \nonumber \\ &+\sum_{l=0}^{m-1}\sum_{k\in[M]\setminus D}\mathbf{A}_l\mathbf{\tilde B}_k(x_{j_k})x^l+\sum_{k\in [M]\setminus D}{\mathbf{R}\mathbf{\tilde B}_k(x_{j_k})}x^m&&\nonumber \\
&=\sum_{l=0}^{n(m+1)-1}\mathbf{Z}_l x^l,&&\nonumber
\end{flalign}
where $\{\mathbf{Z}_{l}\}_{l=0}^{n(m+1)-1}$ are given by
\begin{gather}
\mathbf{Z}_{l}=\sum_{k\in[M]\setminus D}\mathbf{A}_l\mathbf{\tilde B}_k(x_{j_k}) \:\:\: \forall l\in[0:m-1], \nonumber \\ 
\mathbf{Z}_{l}=\sum_{k\in[M]\setminus D}\mathbf{R}\mathbf{\tilde B}_k(x_{j_k}) \:\:\: \forall l=m, \nonumber \\ 
\mathbf{Z}_{l}=\mathbf{R}\mathbf{B}_{D,l} \:\:\: \forall l=m+p(m+1), p\in[1:n-1],  \nonumber \\ 
\mathbf{Z}_{l}=\mathbf{A}_{l-p(m+1)}\mathbf{B}_{D,p} \:\:\: \forall l\in[p(m+1):p(m+1)+m-1], \nonumber \\ p\in[1:n-1]. \nonumber 
\end{gather}

Since the degree of polynomial $\mathbf{\tilde A}(x_{i})(\mathbf{\tilde B}_D(x_i)+\sum_{k\in [M]\setminus D}{\mathbf{\tilde B}_k(x_{j_k})})$ is $mn+n-1$ and the evaluating points $\{x_{i}\}_{i=1}^{N}$ are distinct from each other, the master can decode the polynomial from the sub-computation results returned by the $mn+n$ fastest workers, by polynomial interpolation.
We denote the decoding function by $d_D$ and sub-computation result returned from the $i$th fastest worker by $S_{i}$.
The master can decode the polynomial $\mathbf{\tilde A}(x_{i})(\mathbf{\tilde B}_D(x_i)+\sum_{k\in [M]\setminus D}{\mathbf{\tilde B}_k(x_{j_k})})$ from $\{S_{i}\}_{i=1}^{mn+n}$, so that the coefficients $\{\mathbf{Z}_{l}\}_{l=0}^{mn+n-1}$ are obtained.
The whole computation $\mathbf{AB}_D$ can be recovered from the coefficients $\{\mathbf{Z}_l|l\in[p(m+1):p(m+1)+m-1],p\in[1:n-1]\}$.
Therefore, the recovery threshold $K$ equals $mn+n$ and  $d_D(S_{1}, S_{2}, \cdots, S_{mn+n})=\mathbf{AB}_D$.

\begin{remk}
In the private polynomial codes in \cite{Arxiv}, the master's own data $\mathbf{A}$ and the library $\mathbf{B}$ are encoded into separate polynomials.
That is, $\mathbf{\tilde A}$ is polynomial of $x$ whereas $\{\mathbf{\tilde B}_k\}_{k=1}^M$ are polynomials of $y$.
Since the random matrix $\mathbf{R}$ is not considered in the private polynomial codes, $\mathbf{A}$ and $\mathbf{B}$ should not be encoded with same variable $x$ for protecting the master's privacy.
If $\mathbf{A}$ and $\mathbf{B}$ are encoded with same variable $x$ without $\mathbf{R}$, the workers obtain non-zero information about $\mathbf{\tilde A}(x_i)$.
That is, the workers may identify that $\mathbf{A}$ is encoded with $x_i$.
Since the desired matrix $\mathbf{B}_D$ is also encoded with $x_i$, the workers thereby realize that $\mathbf{B}_D$ is desired by the master, thus implying that the master's privacy is violated.
Therefore, compared to the private polynomial codes in \cite{Arxiv}, our private secure polynomial codes have a notable difference.
\label{remk1}
\end{remk}

\subsection{Privacy and Security Proof}
\label{proof}

To prove that the data security and the master's privacy are protected in the private secure polynomial codes, we need to show that the privacy constraint in (\ref{privacyconstraint}) and the data security constraint in (\ref{securityconstraint}) are satisfied for every worker.
The basic idea for proof is similar to that in \cite{Arxiv}.
The details of the proof will be given in Appendix.

\section{Simulation results}
In this section, in terms of the computation time consumed for receiving $K$ sub-computation results across $N$ workers, we compare the private secure polynomial codes and private polynomial codes modified for private secure coded computation. 
In private polynomial codes, the data security can be protected by adding $\mathbf{R}$ when encoding $\mathbf{A}$, which is same as private secure polynomial codes.
Nevertheless, as explained in Remark \ref{remk1}, since the library $\mathbf{B}$ is encoded with distinct variable $y$, the workers are still divided into groups in private polynomial codes.
Grouping may increase the computation time since the computation time is determined by the slowest group.

We assume that the computation time distribution of each worker is independent of each other and follows the exponential distribution as in \cite{KLee}.
In \cite{KLee}, the computation time is given by 
\begin{align}
t_{conv} = \frac{1}{K}(\gamma + \frac{1}{\mu}\text{log}\frac{N}{N-K}),
\label{conv} 
\end{align} 
where $\mu$ and $\gamma$ are the straggling parameter and the shift parameter, respectively. 

Compared to (\ref{conv}), in private secure polynomial codes, the recovery threshold $K$ equals to $n(m+1)$ and $n/(n-1)$ times more computation than directly computing $\textbf{AB}_D$ is required. 
Therefore, the computation time of private secure polynomial code is given by 
\begin{eqnarray}
\label{tps}
t_{ps}=\frac{1}{(m+1)(n-1)}(\gamma + \frac{1}{\mu}\text{log}\frac{N}{N-n(m+1)}).
\end{eqnarray}
The computation time of private polynomial codes is given by (9) in \cite{Arxiv}.

\begin{figure}[t]
    \centerline{\includegraphics[width=7.0cm]{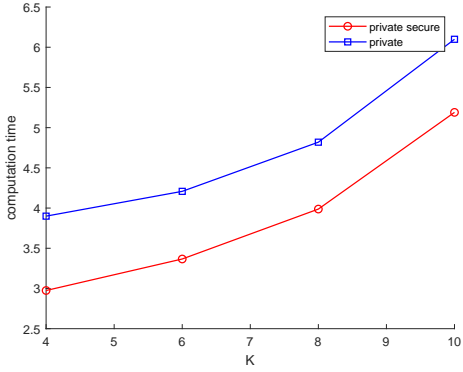}}
    \caption{The computation time comparison between the private secure polynomial code and private polynomial code for $N=12$, $M=4$, $\mu=\gamma=0.1$, and varying $K$.}
    \label{computation}
    \vspace{0mm}
\end{figure}

We compare the computation time between three schemes for $N=12$, $M=4$, and $\gamma=\mu=0.1$. 
We set $n=2$, which implies that the workers are divided into 2 groups in private polynomial codes. 
For fair comparison, we assume that each worker returns only one sub-computation result to the master in private polynomial codes, as in private secure polynomial codes.
Since $K=n(m+1)$ in the private secure polynomial codes and private polynomial codes, we set $K$ as even number and vary $K$ from $4$ to $10$.
That is, $m$ is varying from 1 to 4.
The comparison result for computation time is given in Fig. \ref{computation}. 
Compared to the private polynomial codes, private secure polynomial codes achieves at most $25\%$ reduction in computation time, thus implying that our proposed scheme outperforms the previous works when considering both of data security and master's privacy.

\begin{figure}[t]
    \centerline{\includegraphics[width=7.0cm]{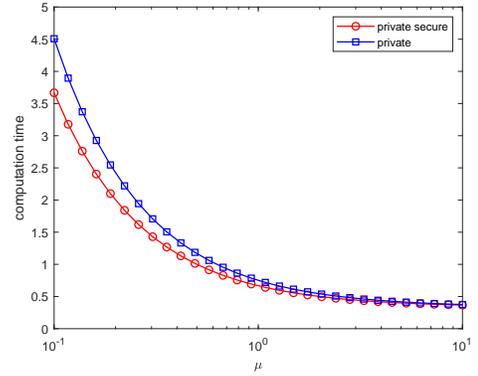}}
    \caption{The computation time comparison between the private secure polynomial code and private polynomial code for $N=12$, $M=4$, $K=4$, 
    $\gamma=1$, and varying $\mu$.}
    \label{computation2}
    \vspace{0mm}
\end{figure}

We also compare the computation time for given $N,M,K,\gamma$ and varying $\mu$. 
We set $N=12$, $M=4$, $K=4$, $\gamma=1$ and vary $\mu$ from $10^{-1}$ to $10$.
As seen in Fig. \ref{computation2}, the private secure polynomial codes outperforms the private polynomial codes, whereas the gap between the two schemes decrease as $\mu$ becomes larger.
This is because the effect of stragglers are diminished when $\mu$ becomes larger.

\section{Conclusion}
In this paper, we introduced private secure coded computation as a variation of coded computation that ensures data security and master's privacy at the same time.
As an achievable scheme for private secure coded computation, we proposed private secure polynomial codes based on private polynomial codes in private coded computation. 
In simulations, we compared private secure polynomial codes with private polynomial codes in terms of computation time, and showed that the proposed scheme outperforms the existing scheme.  
In future work, we will further analyze the performance of private secure polynomial codes and compare the performance in practical scenarios, e.g., AWS or Google Cloud.

\section{Appendix}
In this appendix, we prove that the private secure polynomial codes ensure both of the data security and the master's privacy.
In other words, we show that the data security constraint in (\ref{privacyconstraint}) and the privacy constraint in (\ref{securityconstraint}) are satisfied for every worker.

For the master's privacy, we need to show that the privacy constraint for each worker $\text{W}_i$ is satisfied, which was given by (\ref{privacyconstraint}).
By chain rule, we can write the privacy constraint as follows.
\begin{flalign}
&I(D;Q_i^D,C_i^D,R_i^D,\mathbf{B}) \nonumber\\
&=I(D;Q_i^D) \nonumber\\
&+I(D;\mathbf{B}|Q_i^D) \nonumber\\
&+I(D;C_i^D|Q_i^D,\mathbf{B}) \nonumber \\
&+I(D;R_i^D|Q_i^D,\mathbf{B},C_i^D) \nonumber
\end{flalign}

Note that $R_i^D=f^D_{\text{W}_i}(g^D_{\mathbf{A},\text{W}_i}(\mathbf{A}), g^D_{\mathbf{B},\text{W}_i}(\mathbf{B}))$ is a deterministic function of $C_i^D=g^D_{\mathbf{A},\text{W}_i}(\mathbf{A})$ and $g^D_{\mathbf{B},\text{W}_i}(\mathbf{B})$, where $g^D_{\mathbf{B},\text{W}_i}(\mathbf{B})$ is a function of $\mathbf{B}$. 
Since $g^D_{\mathbf{B},\text{W}_i}(\mathbf{B})=\mathbf{\tilde B}_D(x_{i})+\sum_{k\in [M]\setminus D}{\mathbf{\tilde B}_k(x_{j_k})}$ and the evaluating points $x_i$ and $\{x_{j_k}|k\in[M]\setminus D\}$ are determined by the the queries $Q_i^D$, $g^D_{\mathbf{B},\text{W}_i}(\mathbf{B})$ is a deterministic function of $Q_i^D$, which implies that $R_i^D$ is a deterministic function of $C_i^D$, $\mathbf{B}$, and $Q_i^D$.
Therefore, $I(D;R_i^D|Q_i^D,\mathbf{B},C_i^D)=0$. 
Since $C_i^D=\mathbf{\tilde A}(x_{i})=\sum_{l=0}^{m-1}{\mathbf{A}_lx_{i}^l}$, $C_i^D$ is independent of $D$. 
Therefore, $I(D;C_i^D|Q_i^D,\mathbf{B})=0$. 
Since the master determines the index of the desired matrix $D$ without knowing any information of the library $\mathbf{B}$, the library $\mathbf{B}$ is independent of $D$, which is followed by $I(D;\mathbf{B}|Q_i^D)=0$. 

For the desired matrix $\mathbf{B}_{D}$, the master sends queries $Q_i^D$ to each worker $\text{W}_i$ in order to request $\text{W}_i$ to encode the library $\mathbf{B}$ into $\mathbf{\tilde B}_D(x_i)+\sum_{k\in [M]\setminus D}{\mathbf{\tilde B}_k(x_{j_k})}$ and compute the function $\mathbf{\tilde A}(x_i)(\mathbf{\tilde B}_D(x_i)+\sum_{k\in [M]\setminus D}{\mathbf{\tilde B}_k(x_{j_k})})$.
The queries $Q_i^D$ are fourfold:
\begin{enumerate}
    \item $Q_{i,p}^D$ : queries for partitioning each matrix $\mathbf{B}_{k}$ in the library $\mathbf{B}$ into $n-1$ submatrices $\{\mathbf{B}_{k,l}\}_{l=1}^{n-1}$
    \item $Q_{i,e}^D$ : queries for evaluating $\mathbf{\tilde B}_{D}$ and $\{\mathbf{\tilde B}_{k}|k\in [M]\setminus D\}$ at the points $x_i$ and $\{x_{j_k}|k\in[M]\setminus D\}$, respectively
    \item $Q_{i,s}^D$ : queries for summing the evaluations of $\{\mathbf{\tilde B}_{k}\}_{k=1}^M$ into one equation $\mathbf{\tilde B}_D(x_i)+\sum_{k\in [M]\setminus D}{\mathbf{\tilde B}_k(x_{j_k})}$
    \item $Q_{i,c}^D$ : queries for computing the function
\end{enumerate}

According to $Q_{i,p}^D$, all submatrices $\{\mathbf{B}_{k,l}\}_{(k,l)=(1,1)}^{(M,n-1)}$ are elements in $\mathbb{F}_q^{s\times t/(n-1)}$.
Therefore, $Q_{i,p}^D$ are independent of $D$, which implies $I(D;Q_{i,p}^D)=0$.

According to $Q_{i,e}^D$, as assumed in Section \ref{general}, the points $x_i$ and $\{x_{j_k} | k\in [M]\setminus D\}$ are distinct from each other and randomly chosen in $\mathbb{F}_q$.
Therefore, $Q_{i,e}^D$ are independent of $D$, which implies $I(D;Q_{i,e}^D)=0$.

According to $Q_{i,s}^D$, all of the evaluations of $\{\mathbf{\tilde B}_k\}_{k=1}^M$ are symmetrically summed into one equation $\mathbf{\tilde B}_D(x_i)+\sum_{k\in [M]\setminus D}{\mathbf{\tilde B}_k(x_{j_k})}$.
Therefore, $Q_{i,s}^D$ are independent of $D$, which implies $I(D;Q_{i,s}^D)=0$.

According to $Q_{i,c}^D$, $C_i^D=\mathbf{\tilde A}(x_{i})$ is multiplied by $\mathbf{\tilde B}_D(x_i)+\sum_{k\in [M]\setminus D}{\mathbf{\tilde B}_k(x_{j_k})}$.
We already explained that $C_i^D$ and $\mathbf{\tilde B}_D(x_i)+\sum_{k\in [M]\setminus D}{\mathbf{\tilde B}_k(x_{j_k})}$ are independent of $D$.
Therefore, $Q_{i,c}^D$ are also independent of $D$, which implies $I(D;Q_{i,c}^D)=0$.

As a result, $I(D;Q_{i}^D)=0$, which implies that $I(D;Q_i^D,C_i^D,R_i^D,\mathbf{B})=0$.
Since the privacy constraint is satisfied for every worker, the master's privacy is considered to be protected in private  polynomial codes. $\square$

For the data security, we need to show that the data security constraint given in (\ref{securityconstraint}).
By the chain rule, we can write the data security constraint as follows.
\begin{flalign}
&I(\mathbf{A};Q_i^D,C_i^D,R_i^D,\mathbf{B}) \nonumber\\
&=I(\mathbf{A};Q_i^D) \nonumber\\
&+I(\mathbf{A};\mathbf{B}|Q_i^D) \nonumber\\
&+I(\mathbf{A};C_i^D|Q_i^D,\mathbf{B}) \nonumber \\
&+I(\mathbf{A};R_i^D|Q_i^D,\mathbf{B},C_i^D) \nonumber
\end{flalign}

As explained, $R_i^D$ is a deterministic function of $C_i^D$, $\mathbf{B}$, and $Q_i^D$.
Therefore, $I(\mathbf{A};R_i^D|Q_i^D,\mathbf{B},C_i^D)=0$. 
Since $C_i^D=\mathbf{A}_l x_i^{l}+\mathbf{R}x_i^m$ and $\mathbf{R}$ is a random matrix, $I(\mathbf{A};C_i^D|Q_i^D,\mathbf{B})$.
Since $\mathbf{A}$ and $\mathbf{B}$ are independent, $I(\mathbf{A};\mathbf{B}|Q_i^D)=0$.
Since $Q_i^D$ are queries for encoding $\mathbf{B}$, $Q_i^D$ are independent of $\mathbf{A}$, thus implying that $I(\mathbf{A};Q_i^D)$.
As a result, $I(\mathbf{A};Q_i^D,C_i^D,R_i^D,\mathbf{B})=0$.
Since the security constraint is satisfied for every worker, the data security is ensured in private secure polynomial codes. $\square$

\end{document}